\documentclass[final,5p,times,twocolumn]{elsarticle}
\usepackage{amsmath}
\usepackage{hyperref}
\usepackage{cleveref}

\usepackage{amssymb}
\usepackage{amsthm}
\usepackage{textcomp}
\usepackage{textgreek}
\usepackage{lineno}

\journal{Nuclear Instruments and Methods in Physics Research A}

\begin{document}
 
\begin{frontmatter}



\title{Wake fields effects in dielectric capillary}

\author[label1]{A. Biagioni}
\author[label1]{M.P. Anania}
\author[label1]{M. Bellaveglia}
\author[label1]{E. Brentegani}
\author[label1]{G. Castorina}
\author[label1]{E. Chiadroni}
\author[label2]{A. Cianchi}
\author[label1]{D. Di Giovenale}
\author[label1]{G. Di Pirro}
\author[label5]{H. Fares}
\author[label3]{L. Ficcadenti}
\author[label1]{F. Filippi}
\author[label1]{M. Ferrario}
\author[label3]{A. Mostacci}
\author[label1]{R. Pompili}
\author[label1]{J. Scifo}
\author[label1]{B. Spataro}
\author[label1]{C. Vaccarezza}
\author[label1]{F. Villa}
\author[label4]{A. Zigler}

\address[label1]{Laboratori Nazionali di Frascati, INFN, Via E. Fermi 40, 00044 Frascati, Italia}
\address[label2]{Dipartimento di Fisica, Universit\`{a} di Roma Tor Vergata, V. della Ricerca Scientifica 1, 00133 Roma, Italia}
\address[label3]{Dipartimento di Scienze di Base e Applicate per l'Ingegneria (SBAI), Sapienza Universit\`{a} di Roma,
Via A. Scarpa 14-16, 00161 Roma, Italia}
\address[label4]{Hebrew University of Jerusalem, Jerusalem 91904 (Israel)}
\address[label5]{Assiut University, Department of Physics, Faculty of Science, Assiut 71516, Egypt}


\address{}

\begin{abstract}
Plasma wake-field acceleration experiments are performed at the SPARC\_LAB test facility by using a gas-filled capillary plasma source composed of a dielectric capillary. The electron can reach GeV energy in a few centimeters, with an accelerating gradient orders of magnitude larger than provided by conventional techniques. In this acceleration scheme, wake fields produced by passing electron beams through dielectric structures can determine a strong beam instability that represents an important hurdle towards the capability to focus high-current electron beams in the transverse plane. For these reasons, the estimation of the transverse wake-field amplitudes assumes a fundamental role in the implementation of the plasma wake-field acceleration. In this work, it presented a study to investigate which parameters affect the wake-field formation inside a cylindrical dielectric structure, both the capillary dimensions and the beam parameters, and it is introduced a quantitative evaluation of the longitudinal and transverse electric fields.

\end{abstract}


\end{frontmatter}



\section{Introduction}
\label{}

We are interested in the estimation of wake fields created inside a dielectric capillary by a single bunch in order to determine how such electric fields affect the stability of the bunch itself. For this reason, it is not taken into account the wake-field propagation within the capillary towards any other witness bunch trailing the drive bunch, which is what it has to be made to implement the dielectric wakefield acceleration. Hence, our study concerns the analysis of the wake fields acting on a single bunch that passes through the gas-filled dielectric capillary, which are created by the bunch's head and act on the bunch itself, especially to the tail. In general, the electron bunch propagation along a dielectric structure causes the excitation of high-amplitude $TM_{mn}$ waveguide modes leading to formation of wake fields, which move in synchronism with the bunch because they have phase velocities equal to the bunch velocity, that is close to $c$. For this study, TE-modes ($E_z$ = 0 and $H_z\neq  0$) can be ignored since they cannot accelerate charged particles along $z$-direction within a dielectric-loaded circular waveguide, therefore we will concentrate on the accelerating TM-modes ($E_z\neq 0$ and $H_z$ = 0) since they can be excited by the charged particles \cite{label1,label2}. With regard to the waveguide modes, the subscript $m$ gives the azimuthal mode order (it determines which term has to be considered in the multipole expansion: for istance, the monopole term correspond to $m$ = 0, and so on) and the subscript $n$ gives the radial modes that correspond to the eigenfrequencies of such azimuthal mode.
In general, the intensity of the wake-field is directly proportional to the drive bunch charge $Q$ and it is inversely proportional to the transverse dimensions of the dielectric capillary. Also, the bunch length $\sigma_z$ strongly affects the amplitude of wake fields: the shorter is the bunch length, the higher is the wake-field along the capillary \cite{label3}.

\section{Experimental measurements}
\label{}
The capillary used in our experimental set-up is a dielectric device that is composed of two different sections (Fig. 1).
\begin{figure}[ht]
	\centering
	\includegraphics[scale=0.51,trim=0cm 0cm 0cm 0cm]{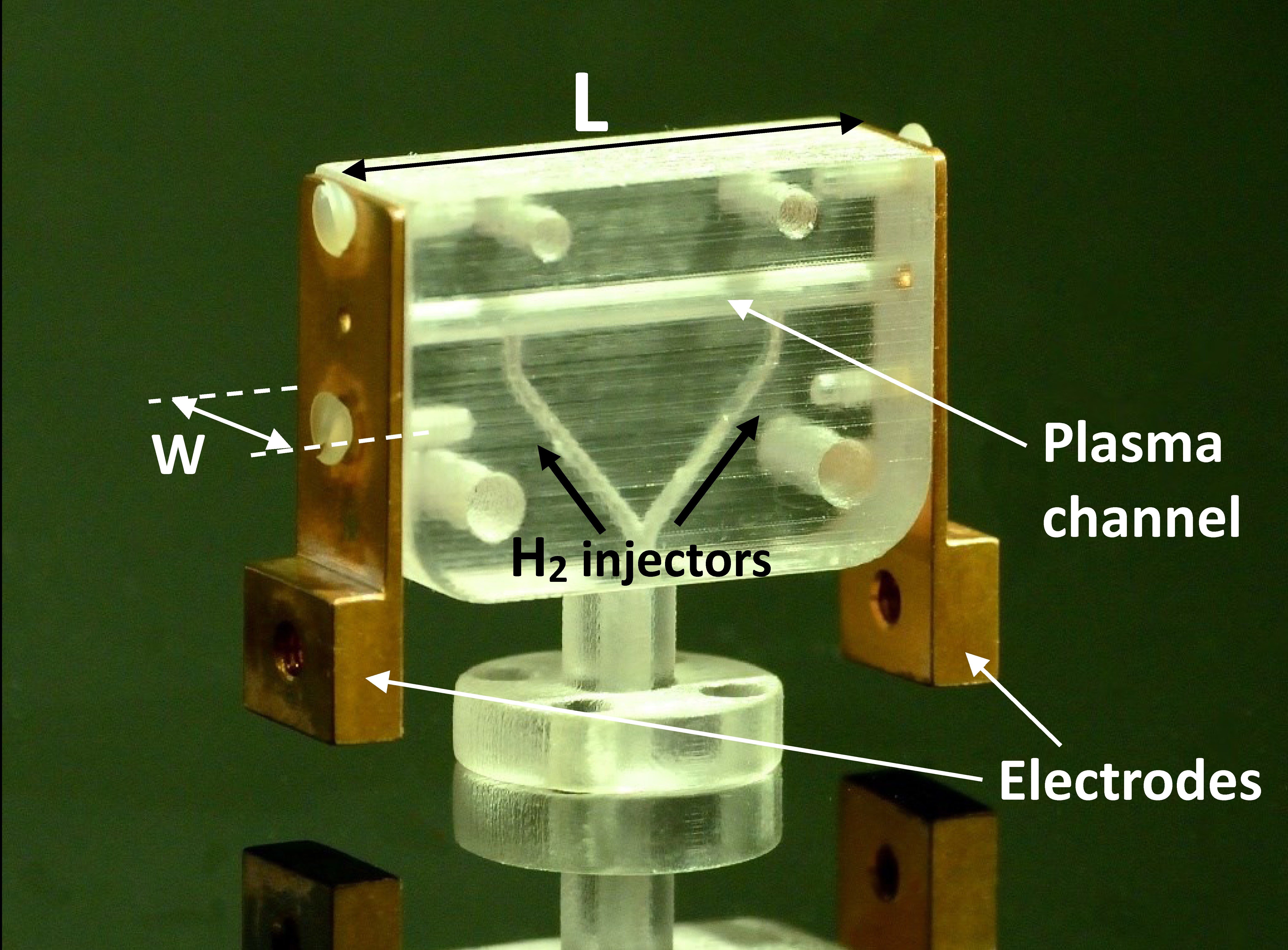}
	\caption{Picture of the dielectric capillary plasma source used at SPARC\_LAB test facility to analyze the wake fields effects on electron beams passing through such structures.}
\end{figure}

\begin{figure*}
	\centering
	\includegraphics[scale=0.18,trim=0cm 0cm 0cm 0cm]{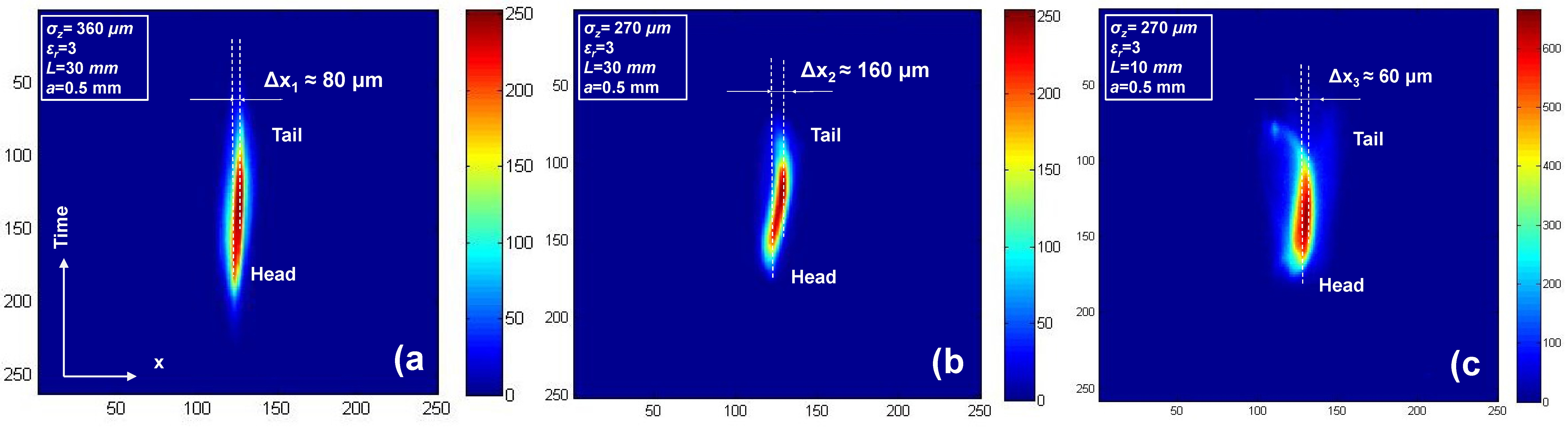}
	\caption{Experimental measurements of wake fields effects on electron bunches at different operating conditions both beam parameters and geometric features of the capillary.}
\end{figure*}
The first one is a backing section, which is made of a transparent plastic material \cite{label4,label5,label6,label7} and built by using a 3D printing device. The second section, which includes the plasma channel, is separately created in a glass cylinder of around 3 mm in radius (a dielectric material with similar relative permittivity of the backing section) to prevent any damage caused by the electron beam. This cylinder is then inserted into the backing part. The plasma channel inside the glass cylinder has a circular cross section of 0.5 mm in radius and fed by two gas injectors of 300 $\mu$m in radius. We have considered such a structure as a single piece of dielectric material. Finally, two electrodes at the ends of the capillary provide the high voltage to the hydrogen gas.
The capillary dimensions considered in our experiments are as follows: two different channel lengths of 30 and 10 mm, for which the inner radius is $a$ = 0.5 mm and the relative permittivity is $\epsilon_r$ = 3. With regard to the outer radius $b$ of the capillary, since the plasma channel is created in a plastic holder which is 8 mm wide, we assume that it is 4 mm in radius.
Several experimental conditions have been considered in order to investigate the wake fields production as a function of both the beam parameters and the geometric features. Figure 2 shows some effects produced by wake fields created inside our capillaries of 30 and 10 mm length, when the electron bunches have two longitudinal lenghts of 360 $\mu$m and 270 $\mu$m. The beam charge is $Q$ = 50 pC, in both cases.
A first comparison can be made between two electron beams passing through the same capillary, which have 360 $\mu$m and 270 $\mu$m of the longitudinal length, as shown in Figure 2a and Figure 2b. In these conditions, the tail deflection with respect to the head is $\Delta$$x_1$ $\approx$ 80 $\mu$m for the longer bunch, but it becames $\Delta$$x_2$ $\approx$ 160 $\mu$m for the shorter one. A second comparison concerns electron bunches with same longitudinal lengths, $\sigma_z$ = 270 $\mu$m, that pass through two different plasma channel of 10 and 30 mm length, as shown in Figure 2c and Figure 2b respectively. The tail displacement we obtain if the bunch crosses a capillary of 10 mm length is $\Delta$$x_3$ $\approx$ 60 $\mu$m, that has to be compared to $\Delta$$x_2$, which is related to a capillary of 30 mm length. It should be noted that all frames reported in Figure 2, for each operating condition, correspond to the maximum value, over 100 shots, of the tail deflection we have measured; in fact, the transverse movement of bunch tails can change depending on the bunch's position inside the capillary.
Such results highlight that the shorter is the bunch, the larger is the amplitude of wake fields produced by electron beams passing through a dielectric structure. Also, the longer is the capillary, the larger is the wakefield's effect on the bunch.

\section{Theoretical model}
\label{}
A general equation of the electric field, transverse or longitudinal, produced when a bunch of charge passes through a cylindrical dielectric capillary with speed $c$ and an offset $r_0$ from the center axis, can be expressed as \cite{label8,label9}
\begin{equation}
\label{eqn:generalequation}
E=A\sum_{m=0}^{m=\infty}\sum_{n=0}^{n=\infty}B_{mn}(a,b,r_0,\sigma_z,\epsilon_r,\omega_r)cos(\frac{\omega_{mn}}{c}z_0),
\end{equation}
 where $B_{mn}$ depends on the Bessel functions of the first and the second kind and \emph{A} is proportional to the beam charge; also, the parameter $\omega_{mn}$ represents the eigenfrequencies of \emph{TM}$_{mn}$ modes. If we know the electric field within the dielectric structure, the wake potential can be defined as follow \cite{label3}: 
\begin{equation}
\label{eqn:wakepotentialequation}
W=-e\int_{0}^{L} Edz,
\end{equation}
where $L$ is the length of the capillary. The analysis of the beam instabilities caused by wake fields in dielectric structures has to start from the calculation of the longitudinal electric field, which leads to the transverse electric field only if the bunch is misaligned with offset $r_0$ from the center axis. On the contrary, the longitudinal electric field is always created by the bunch, whether it is on-axis or out-of axis, and it represents a deceleration term \cite{label10,label11}.  
Therefore, the transverse electric field can be evaluated by Panofsky-Wenzel theorem \cite{label12,label13}:
\begin{equation}
\label{eqn:PanofskyWenzelequation}
q\nabla_t E_z=\frac{\partial F_t}{\partial z},
\end{equation}
This transverse force $F_t$ represents the source of the head-tail single bunch beam instabilities within a dielectric structure. In fact, when a moving charge passes through the capillary, the energy loss of the head produces two different effects: a longitudinal wake-field which is equivalent to application of an effective drag force on the bunch, that is produced in every condition, and a transverse wake-field that acts on the tail producing a transverse movement of it with respect to the head, only if there is an off-axis displacement of the beam.
\begin{figure}[ht]
\centering
\includegraphics[scale=0.6,trim=0cm 0cm 0cm 0cm]{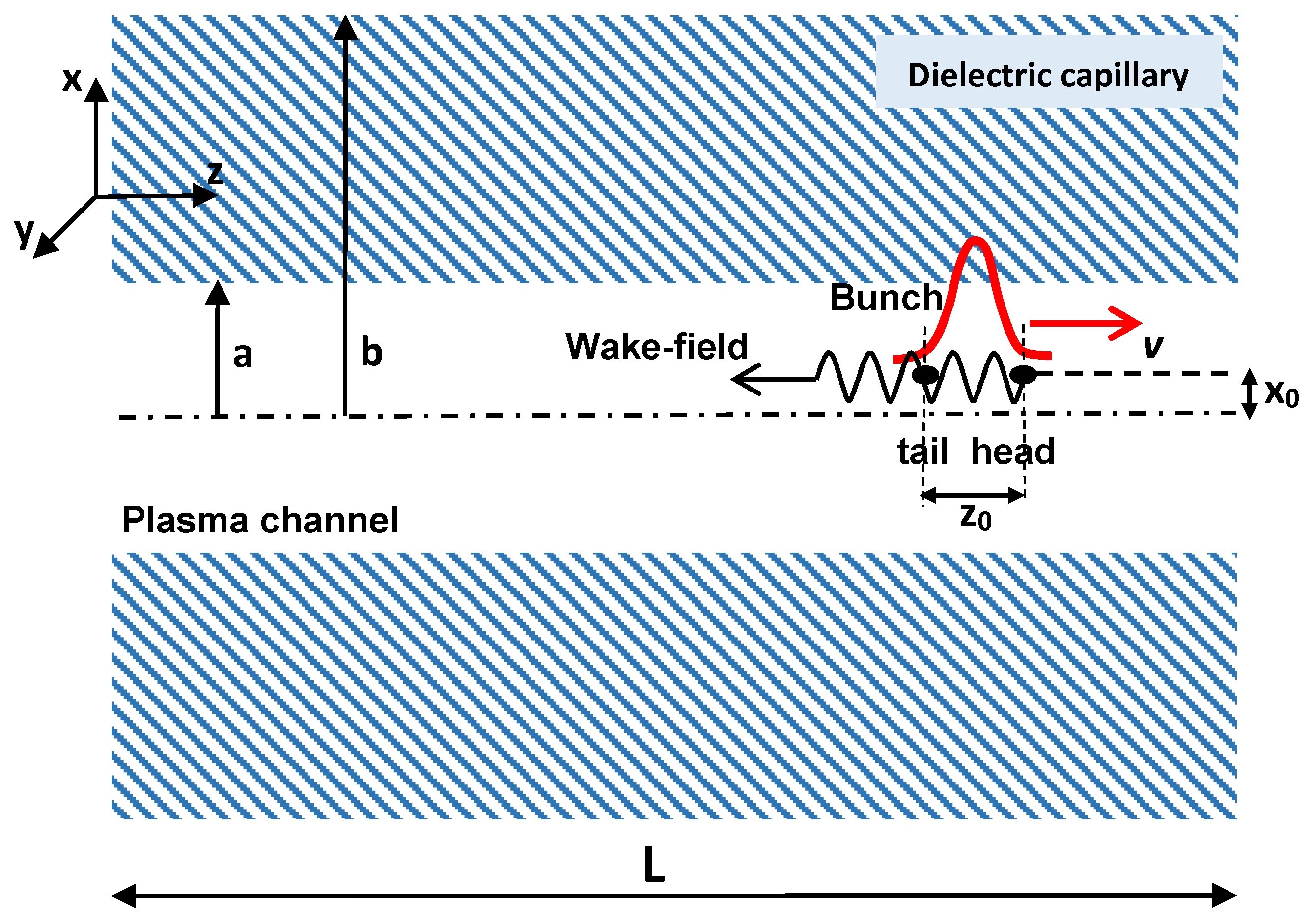} 
\caption{Sketch of a 2D geometry that describes the wake-field structure corresponding to a plasma capillary source: $a$ = 0.5 mm, $b$ = 4 mm while $L$ has been chosen of 10 and 30 mm length.}
\end{figure}
So far, it has been given a general description of the waveguide modes which are responsible of the wake fields inside our dielectric structure. In this work, with a view to simplifying the problem analysis, we are going to study the wake-field generation by using a simplified 2D geometry, for which we have considered two parallel layers of dielectric material to simulate the capillary. We have considered, as a source for the wake fields, a bunch of charge $Q$ moving at the speed of light $c$ in the plasma channel (without plasma). The direction of motion of the bunch is along the $z$-axis, but it is misaligned with off-set $x_0$ from the center axis. The distance between the head and the tail of bunch is $z_0=z-vt$. By using the 2D geometry, compared to the 3D one, we can ignore the azimuthal order $m$, because this parameter takes into account the cylindrical symmetry of the structure. Figure 3 provides a schematic description of the planar wake-field structure.
The theoretical model presented in this paper is related to an ultrarelativistic single bunch that travels inside the plasma channel with $\beta$ $\simeq$1 and $\gamma$ = 250. Furthermore, it has been considered a total charge $Q$ = 50 pC and a Gaussian longitudinal profile of the e-bunch, as well as two different bunch lengths of 270 and 360 $\mu$m, in order to be congruent with our experimental results. 
For this kind of dielectric waveguide, the longitudinal and transverse electric field can be written as \cite{label14}:

\begin{equation}
\label{eqn:Ezequation}
E_z(x,z,t)=
-\frac{Q}{2a\epsilon_r\epsilon_0}\sum_{n}\left[\frac{f_n(x)}{\alpha_n}\right]exp\left[-\frac{(\omega_n\sigma_z)^2}{4c^2}\right] exp\left[-i\frac{\omega_n}{c}z_0\right]
\end{equation}

\begin{equation}
\label{eqn:Exequation}
E_x(x,z,t)=
 -i\frac{Q}{2a\epsilon_r\epsilon_0}\sum_{n}\left[\frac{g_n(x)}{\alpha_n}\right]exp\left[-\frac{(\omega_n\sigma_z)^2}{4c^2}\right] exp\left[-i\frac{\omega_n}{c}z_0\right]. 
\end{equation}

It should be noted that, for a 3D geometry, these equations should be a functions of the azimuthal order $m$, but, given that we have considered a 2D geometry, the electric field is not related to the azimuthal order (θ angle of the cylindrical coordinates), but it only depends on the eigenfrequencies $\omega_n$ of the general TM-mode $m$. The expressions of the parameters $f_n$(x) and $g_n$(x) are written below:

\begin{equation}
\begin{cases}
\label{eqn:fngnequation}
f_n(x)=cosh(k_nx) & \text{se $-a \leq x \leq a$}\\g_n(x)=\gamma sinh(k_nx) & \text{se $-a \leq x \leq a$},
\end{cases}
\end{equation}
where $k_n=\omega_n/c\beta\gamma$. All eigenfrequencies $\omega_n$ of waveguide modes can be obtained by calculation of zeros of the conditional equation $C(\omega_n)$:
\begin{equation}
\begin{cases}
\label{eqn:conditionalequation}
C(\omega_n)=\epsilon_r k_ncot(p_n(b-a))\\p_n=\gamma k_n \sqrt{\epsilon_r\beta^2-1}.
\end{cases}
\end{equation}

The longitudinal electric field expressed by Eq. (4) is plotted in Figure 4. The amplitude of the electron bunch has been scaled in order to overlap it with the electric fields.
Figure 4 shows the longitudinal electric field $E_z(x = 0, z_0)$ as a function of the $z$-direction; it is calculated on the $z$-axis ($x$ = 0) and takes into account up to $n$=150 modes that ranging between 13.2 GHz to 3.8 THz; also, it should be noted that the electric field peaks have a typical alternating-sign behaviour.  
\begin{figure}
\centering
\includegraphics[scale=0.45,trim=0cm 0cm 0cm 0cm]{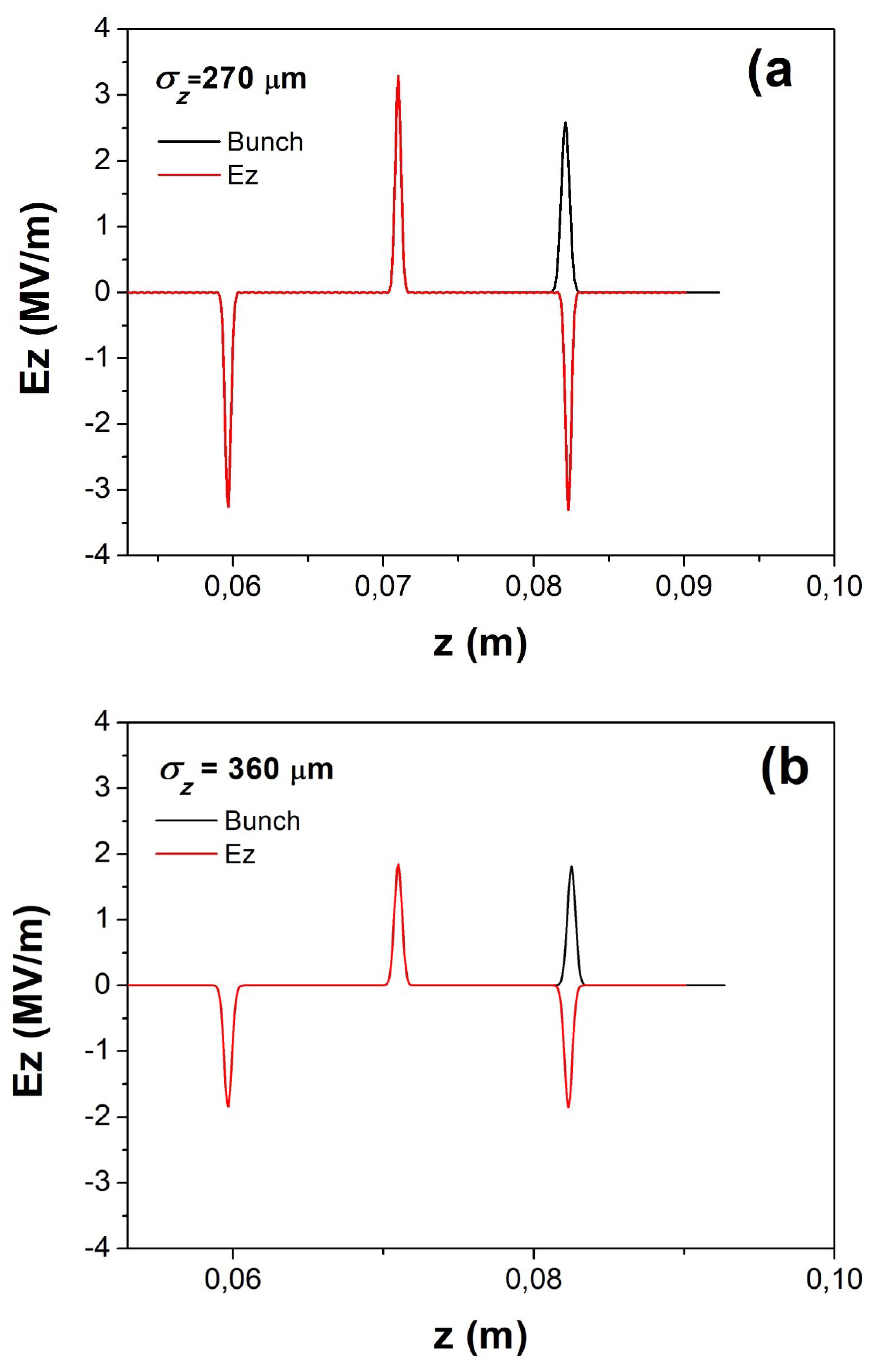}
\caption{Longitudinal electric field obtained if we take into account the first 150 eigenfrequencies; the red curve represents $E_z$ and the black curve is the e-bunch. The entrance of the capillary corresponds to z=0. a) Longitudinal electric field at $\sigma_z$ = 270 $\mu$m; b) Longitudinal electric field at $\sigma_z$ = 360 $\mu$m.}
\end{figure}
Figure 4 also describes a comparison of longitudinal electric fields $E_z$ calculated at two different bunch lengths: $\sigma_z$ = 270 $\mu$m and $\sigma_z$ = 360 $\mu$m, that are the same values we have used to perform our experimental measurements reported in Figure 2. The maximum value of the longitudinal electric field will reach around 3.3 MV/m (Fig. 4a) if the bunch length is 270 $\mu$m and will be around 1.8 MV/m (Fig. 4b) if the bunch is 360 $\mu$m length, which is around half of the first value.
Actually, the formalism expressed by Eqs (4,5) is referred to wake fields created inside a dielectric structure that is surrounded by a perfect-conducting cylinder, while in our capillary there is no any metallic guide. However, we have used such formalism to explain our experimental results, concerning the bunch's displacements, because such equations can describe how the bunch's head creates the electric fields and also the way that they act on the bunch's tail. Therefore, these wake fields created within the e-bunch can determine the tail's movement with respect to the head. Really, the perfect-conducting cylinder around the dielectric structure is used to produce the electric field reflection along the capillary \cite{label8,label9}, from a driver bunch towards other witness bunches trailing the first one (dielectric wake-field acceleration). However, the formalism in Refs [8,9] has been used to give an estimation of the wake fields effects and principally to get a better understanding of the wake fields production trying to study the main parameters that determine such a phenomenon. In our study, therefore, we are only interested in the interaction between a single e-bunch and the first electric field pulse (Fig. 4), which is solely responsible to affect the bunch because they are exactly overlapped. Also, since there is no any perfect-conducting cylinder around the dielectric structure, the other electric pulses do not assume a crucial role to study the wake fields effects within a single bunch.

In order to evaluate the displacement of the bunch tail with respect to the head, actually, we are interested to investigate the transverse electric field $E_x$, that is expressed by Eq. (5) and plotted in Figure 5. This transverse electric field corresponds to a beam off-set $x_0$ = 200 $\mu$m from the center axis. As it has already done for $E_z$, also for $E_x$ we present a comparison relative to two different longitudinal bunch lengths, that are again 270 and 360 $\mu$m: in the first case, the maximum value of the transverse field is around 2.5 MV/m (Fig. 5a) and in the second one it reaches around 1.1 MV/m (Fig.5b), which is, as it happened with $E_z$, around half of the first value. 

\begin{figure}
\centering
\includegraphics[scale=0.45,trim=0cm 0cm 0cm 0cm]{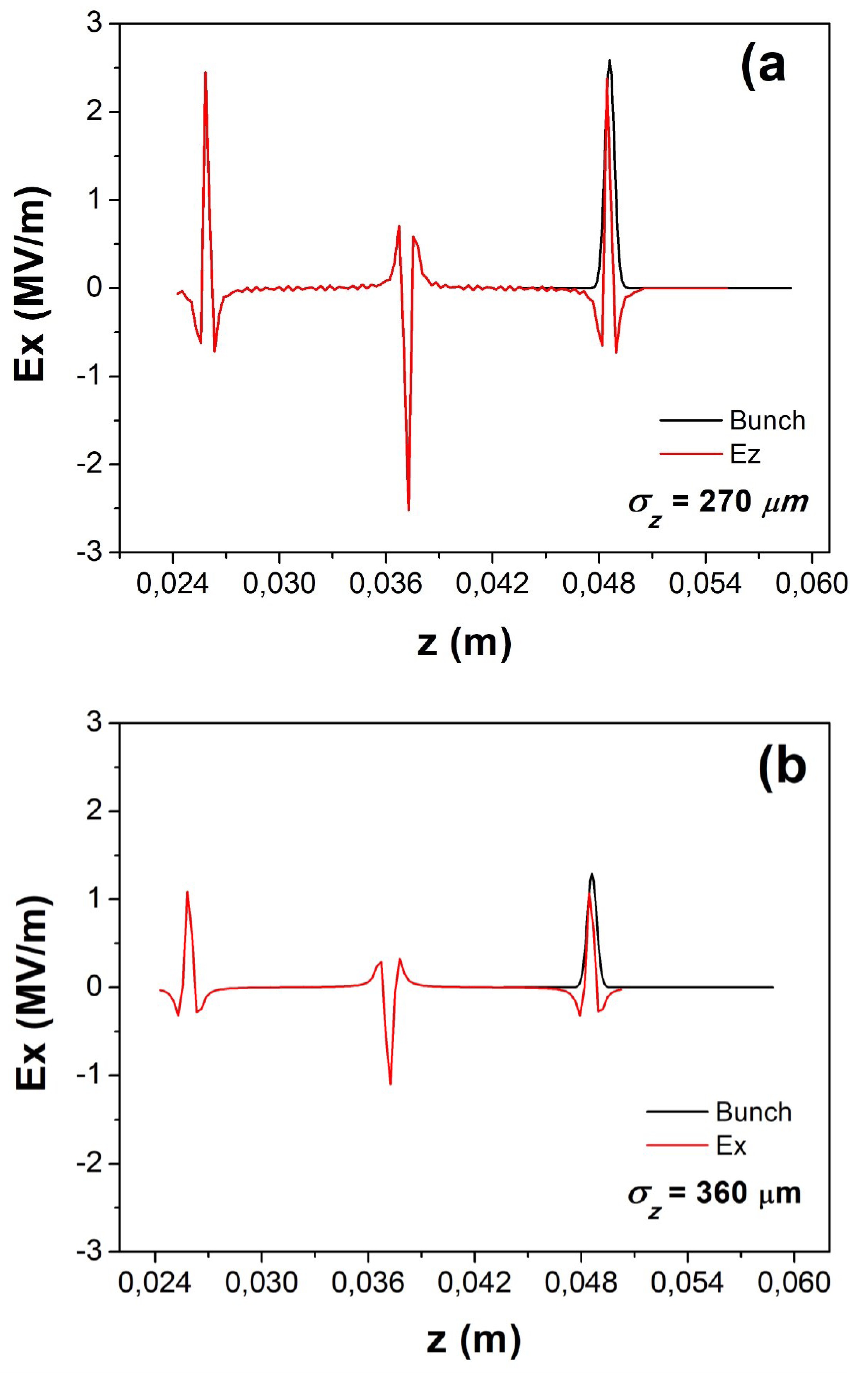}
\caption{Transverse electric field obtained if we take into account the first 150 eigenfrequencies; the red curve represents $E_x$ and the black curve is the e-bunch. The entrance of the capillary corresponds to z=0. a) Transverse electric field at $\sigma_z$ = 270 $\mu$m; b) Longitudinal electric field at $\sigma_z$ = 360 $\mu$m.}
\end{figure}

This result, as regard to the transverse field, is very interesting because also the tails deflections we have measured by using the same operating conditions on beams, maintain a similar ratio (Fig.2): $\Delta x_1$ $\approx$ 160 $\mu$m and $\Delta x_2$ $\approx$ 80 $\mu$m. However, the theoretical model shows that the longer is the bunch, the lower is the electric field produced by bunch itself, both transverse and longitudinal, according to our experimental measurements. Also, it should be noted that electric fields are inversely proportional to the plasma channel radius $a$.     
As previously mentioned, for wake-field studies it is very important to analyze the comparison between the duration time of the electric field pulse and the bunch length. As you can see in the Figures 5, the bunch's head generates the electric field behind itself, therefore it will not be affected by any field; on the contrary, its tail will undergo a transverse displacement, to the right or left depending on the sign of the field, because it follows the head and will be located in a $z$-position where the electric field reach about the maximum value.
  
\begin{figure}
\centering
\includegraphics[scale=0.45,trim=0cm 0cm 0cm 0cm]{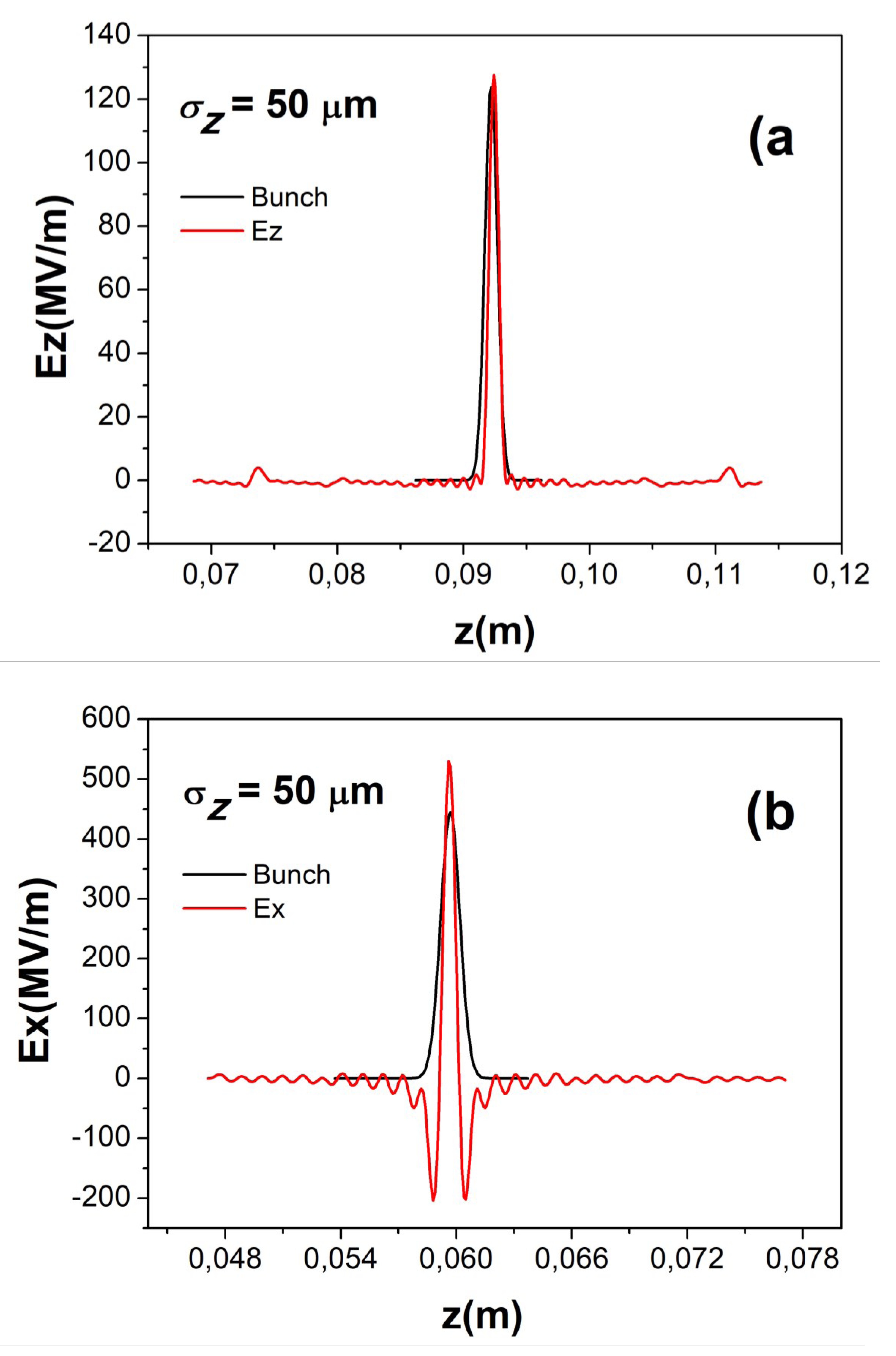}
\caption{Electric fields obtained if we take into account the first 150 eigenfrequencies; the red curve represents $E_x$ and the black curve is the e-bunch. The entrance of the capillary corresponds to z=0. a) Longitudinal electric field at $\sigma_z$ = 50 $\mu$m; b) Transverse electric field at $\sigma_z$ = 50 $\mu$m.}
\end{figure}

Finally, since many of the applications foreseen at the SPARC\_LAB test facility and other plasma-based accelerating machine, as EuPRAXIA \cite{label15} facility or the new INFN infrastructure called EuPRAXIA@SPARC\_LAB \cite{label16}, are going to consider very short electron bunch, few tens of micrometers, we should take into account the wake-field amplitudes that can be produced by using our capillary. With this purpose in mind, other calculations have been performed relative to our dielectric structure, after considering an electron bunch of 50 $\mu$m length. Figure 6 shows longitudinal and transverse electric fields produced in these conditions on beam. The $E_z$ field can reach pick amplitudes around 130 MV/m while the $E_x$ field can arrive up to 550 MV/m when the off-axis displacement $x_0$ is equal to 200 $\mu$m. These values of the wake-field inside a dielectric capillary could produce strong instabilities of the driver bunch, but they can be restored to admissible values by using a capillary that has very different properties as higher values of the radius, according to Eqs. (5) and (6).
\section{Conclusions}
\label{}
Electron beams passing through a dielectric capillary plasma source produce wake fields that determine beam instabilities. This phenomenon depends on different parameters, related to both the geometric properties of the capillary and the beam properties. Experimental measurements show the typical effect produced on electron bunches, that is the tail deflections with respect to the head. Such behaviour has been investigated as regard to a single e-bunch that passes through our dielectric structure, in order to understand which parameters are responsible to affect the beam and to estimate the values of the produced electric fields. The most important parameter that has to be taken into account is the bunch length: longitudinal and transverse electric fields depend on such parameter by means of an exponential function; for this reason, the theoretical model provides low values of the electric fields $E_z$ and $E_x$, around 3.3 and 2.5 MV/m respectively, when the bunch length is 270 $\mu$m, but they would reach very higher values, around 120 and 550 MV/m respectively, if the bunch was 50 $\mu$m length. Several other experimental investigations are needed in order to get a better understanding of the wake fields production within dielectric capillary, as the measurement of the tail displacement at very low values of the bunch length, the study of wake field behaviours at different values of the capillary radius and that one as a function of the off-axis displacement of the beam.                


\section*{ACKNOWLEDGMENTS}
\label{}

This work was supported by the European Unions Horizon
178 2020 research and innovation programme under grant agreement
No. 653782.

One of the authors, H. Fares, would like to acknowledge support
from Academy of Scientific Research and Technology (ASRT) in
Egypt and INFN in Italy (ASRT-INFN joint project).


\section{References}
\label{}



\end{document}